\documentclass[12pt]{article}
\usepackage{amsmath}
\usepackage{amssymb,amsthm}
\usepackage[english]{babel}
\usepackage[textwidth=18.5cm,textheight=22cm]{geometry}

\renewcommand{\d}{\mathrm{d}}
\newcommand{\bt}{\boldsymbol{t}}

\newcommand{\bb}{\boldsymbol{\beta}}

\newtheorem{pro}{Proposition}

\begin{document}

\title{On the singular sector of the Hermitian random matrix model  in the large $N$ limit }

\author{B. Konopelchenko $^{1}$, L. Mart\'{\i}nez Alonso$^{2}$ and E. Medina$^{3}$
\\
\emph{ $^1$ Dipartimento di Fisica, Universit\'a del Salento and
Sezione INFN}
\\ {\emph 73100 Lecce, Italy}\\
\emph{$^2$ Departamento de F\'{\i}sica Te\'orica II, Universidad
Complutense}\\
\emph{E28040 Madrid, Spain}\\
\emph{$^3$ Departamento de Matem\'aticas, Universidad de C\'adiz}\\
\emph{E11510 Puerto Real, C\'adiz, Spain} }

\maketitle \abstract{The singular sector of zero genus case for the
Hermitian random matrix model in the large $N$ limit is analyzed. It is
proved that the singular sector of the hodograph solutions for the underlying
dispersionless Toda hierarchy and the singular sector of the 1-layer Benney
(classical long wave equation) hierarchy are deeply connected. This
property is due to the fact that the hodograph equations for both
hierarchies describe the critical points of solutions of
Euler-Poisson-Darboux equations $E(a,a)$, with $a=-1/2$  for the
dToda  hierarchy and $a=1/2$ for the 1-layer Benney hierarchy.
\vspace*{.5cm}

\begin{center}\begin{minipage}{12cm}
Euler-Poisson-Darboux equation.

\end{minipage}
\end{center}
\newpage

\section{Introduction}

The  Hermitian matrix ($H$) model is nowadays the paradigmatic model in the theory of  random
 matrices ( see e.g. \cite{gin,dei}). The partition function
 for this model is
\begin{equation}\label{0.1}
Z_N=\int \d H \exp\Big(-\dfrac{N}{T}\,\mbox{tr}(\sum_{k\geq
1}t_k\,H^k)\Big),
\end{equation}
where $T>0$ is a real parameter which represents the temperature
and  the integration is performed on the space  of $N\times N$
Hermitian matrices. The large $N$  limit of the $H$ model is of particular
interest since it exhibits many important universality properties.
It turns out that most of the applications of the $H$ model \cite{gin,dei} arise after regularizing the large $N$-limit
solutions at their singular points (double-scaling limit method).
In this sense the analysis and characterization  of the singular
sector of the $H$ model is of great relevance.

 The simplest situation corresponds to the so called \emph{zero
genus case} in which the support of the eigenvalue density reduces
to a single interval $[\beta_1,\beta_2]$ as $N\rightarrow\infty$. Moreover,  it
is well-known ( see e.g.\cite{se,dsl}) that
the endpoints $\bb:=(\beta_1,\beta_2)$  evolve with the temperature $T$ and the coupling constants $ t_k$ according to the dispersionless Toda (dToda)
hierarchy . The first
members of this hierarchy are the dToda equation $v_{TT}=(\log
v)_{t_1t_1}$ or equivalently the dToda system
\begin{equation}\label{dToda}
u_{t_1}=-v_T,\quad v_{t_1}=-v\,u_T,\
\end{equation}
and the 1-layer Benney ($B$) system
\begin{equation}\label{ben}
u_{t_2}=2\,(u\,u_{t_1}+v_{t_1}),\quad v_{t_2}=2\,(u\,v)_{t_1}.
\end{equation}
These systems represent themselves  two distinguished examples of
integrable 2-component hydrodynamical type systems (see e.g.
\cite{dub}). The dToda equation is the 1+1-dimensional version of
the Boyer-Finley equation from the general relativity
\cite{boyer}. It also arises in various problems of fluid
mechanics (see e.g. \cite{min,wieg,kric-wieg}). The $B$ system
describes long waves in shallow water with free surface in a
gravitational field \cite{ben}. It represents the dispersionless
limit of the celebrated nonlinear Schr\"odinger equation
\cite{zak}. Recently, the $B$ system became a crucial ingredient
in the analysis of the universality of critical behaviour for
nonlinear equations \cite{dubgr}. In general, the $B$ system
\eqref{ben} is an excellent laboratory for study properties of
integrable hydrodynamical type systems.

 Thus, the analysis of the singular sector for the $H$ model in the large $N$
 limit reduces to the corresponding
 analysis for the dToda hierarchy. Recently the authors \cite{kmm} have provided
 a method for studying the hodograph solutions and their singular  sectors (gradient catastrophe points)
of the family of coupled KdV hierarchies dcKdV$_m\,$ ($m\geq 1$),  where the
case $m=2$ corresponds to the $B$ hierarchy.  The study of these singular sectors was
already initiated in \cite{kod}.  In the method of \cite{kmm}  the hodograph solutions are described by the critical
points of scalar functions that satisfy Euler-Poisson-Darboux
(EPD) equations \cite{dar}. This property simplifies drastically
the analysis and classification of the singular sectors .

 In the present paper we show that the results of
 \cite{kmm} can be extended to the dToda hierarchy thus providing
 us with complete analysis of the singular sector of $H$ model for the zero genus case. Moreover,  we demonstrate
 that there is deep similarity between the properties of the sets
 of the hodograph solutions for the  dToda and $B$
 hierarchies.

  Our main observation is that in both cases the hodograph solutions represent   the critical
points
\begin{equation}\label{crit}
\dfrac{\partial
W}{\partial\beta_i}=0,\quad i=1,2,
\end{equation}
of a scalar function $W$  which depends linearly on the coordinates $\bt=(t_0,\,t_1,\,t_2,\ldots)$, where $\bt$ is the set of coupling constants  for the $H$ model  ($t_0=T$) and  the set of  flow parameters in the case of the $B$ hierarchy. Moreover, these functions satisfy the Euler-Poisson-Darboux
equation $E(a,a)$ \cite{dar}
\begin{equation}\label{edpa} \everymath{\displaystyle}
(\beta_1\,-\,\beta_2)\, \dfrac{\partial^2
W}{\partial\beta_1\,\partial\beta_2}\,=a\,\Big(\,\dfrac{\partial
W}{\partial\beta_1}\,-\,\dfrac{\partial W}{\partial\beta_2}\Big),
\quad a=\begin{cases}
\quad\dfrac{1}{2} \quad \mbox{for  $W_B$}\\\\
-\dfrac{1}{2}\quad\mbox{ for $W_H$.}
\end{cases}
\end{equation}
Both functions are related according to a transformation \cite{dar} which maps solutions of $E(-1/2,-1/2)$ into solutions $E(1/2,1/2)$
\[
W_B=4\,\dfrac{\partial^2
W_H}{\partial\beta_1\,\partial\beta_2}-t_{H,1},
\]
where  the parameters of both models are identified according to
 \[
 t_{B,n}=(n+2)\,t_{H,n+2},\quad n\geq 0.
\]
The  equation \eqref{edpa} and its multidimensional version are well known for a long time in the classical geometry \cite{dar}.
 Its relevance to the theory of Whitham equations has been demonstrated recently in the papers \cite{kud}-\cite{tian2}.
 The observation made
 in \cite{kmm} that the hodograph equations for the $B$ system \eqref{ben} have the
 form \eqref{crit}-\eqref{edpa}  seems to be new,  though  results close to this have been provided in the papers \cite{dubgr,pav}.

The paper is organized as follows. In Section 2 we show how the hodograph equations of the $B$ system and the $H$ models are the equations for the critical points of certain  solutions of Euler-Poisson-Darboux equations. In Section 3 we use the  Euler-Poisson-Darboux equations to formulate  a common  description of the singular sectors for theses hodograph equations. Our results are summarized into three Propositions which exhibit the deep connection arising  between the singular sectors of both models. Moreover,
the corresponding singular classes are also characterized in terms of the behaviour near $\lambda=\beta_i \,(i=1,2)$ of the $S$-function ($B$ system) and the eigenvalue density $\rho$ ($H$ model). Section 4 is devoted to a method for the explicit determination of singular classes by means of constraints for the $\bt$-parameters. Finally, some illustrative examples are provided.

\section{Hodograph equations and the Euler-Poisson-Darboux equations}

\subsection*{B hierarchy}

The $B$ system \eqref{ben} is a member of a dispersionless integrable hierarchy which describe deformations of the curve
(see e.g. \cite{kod,km}).
\begin{equation}\label{curve}
p^2=(\lambda-\beta_1)\,(\lambda-\beta_2).
\end{equation}
The  flows
$$
\bb(\bt),\quad \bt=(x:=t_0,t_1,t_2,\ldots)
$$
are characterized by the following condition: There exists a family of functions
$S(\lambda,\bt,\bb)$ satisfying
\begin{equation}\label{kdV}
\dfrac{\partial  S(\lambda,\bt,\bb(\bt))}{\partial t_n}=\Omega_n(\lambda,\bb(\bt)),
\quad n\geq 0.
\end{equation}
where
\begin{equation}
\Omega_n(\lambda,\bb)=
\Big(\dfrac{\lambda^{n+1}}{\sqrt{(\lambda-\beta_1)\,(\lambda-\beta_2)}}\Big)_{\oplus}\,\sqrt{(\lambda-\beta_1)\,(\lambda-\beta_2)}.
\end{equation}
Functions $S$ which satisfy \eqref{kdV} are referred to as \emph{action functions}  in the theory of dispersionless integrable systems (see e.g. \cite{kri}). Notice that for $n=0$ Eq.\eqref{kdV} reads
\[
p=\dfrac{\partial S}{\partial x},
\]
so that the system \eqref{kdV} is equivalent to
\begin{equation}\label{kdvp}
\dfrac{\partial p}{\partial t_n}=\,\partial_x\,\Omega_n,
\end{equation}
and, in terms of $\bb$, it can be rewritten in the hydrodynamical form
\begin{equation}\label{eqbeta}
\dfrac{\partial  \beta_i}{\partial t_{n}}=\Big(\Omega_n(\lambda,\bb)\Big)_{\oplus}\Big|_{\lambda=\beta_i}\,\partial_x\,\beta_i,\quad i=1,2.
\end{equation}
The $t_1$-flow of this hierarchy is the $B$ system \eqref{ben} ($t:=t_1$), which in terms of $\bb$ reads
\begin{equation}\label{benr}
\begin{cases}
\partial_{t}\,\beta_1=\dfrac{1}{2}\,(3\,\beta_1+\beta_2)\,\beta_{1\,x},\\ \\
\partial_{t}\,\beta_2=\dfrac{1}{2}\,(3\,\beta_2+\beta_1)\,\beta_{2\,x}.
\end{cases}
\end{equation}

\vspace{0.3cm}
It was proved in \cite{kmm} that the system \eqref{crit} for the function
\begin{equation}\label{w1}
W_B(\bt,\bb)\,:=\,\oint_{\gamma}\dfrac{\d \lambda}{2\,i\,\pi}\,\dfrac{\lambda\,V_B(\lambda,\bt)}{\sqrt{(\lambda-\beta_1)\,(\lambda-\beta_2)}},
\end{equation}
where
\[
V_B(\lambda,\bt)=\sum_{n\geq 0} \lambda^n\,t_{n},
\]
is a system of hodograph equations for the $B$ hierarchy. Moreover, the action function for the corresponding solutions is given by
\begin{equation}\label{sol}
S(\lambda,\bt,\bb)=\sum_{n\geq 0}\, t_{n}\,\Omega_n(\lambda,\bb)=h_B(\lambda,\bt,\bb)\,\sqrt{(\lambda-\beta_1)(\lambda-\beta_2)}.
\end{equation}
where
\[
h_B(\lambda,\bt,\bb):=\Big(\dfrac{\lambda\,V_B(\lambda,\bt)}{\sqrt{(\lambda-\beta_1)(\lambda-\beta_2)}}\Big)_{\oplus}.
\]
Obviously,  $W_B$ satisfies the Euler-Poisson-Darboux equation $E(\frac{1}{2}, \frac{1}{2})$.
Written explicitly, $W_B$ represents itself the series
\begin{align}\label{Wexp}
 W_B&=\,\dfrac{x}{2}(\beta_1+\beta_2)+\dfrac{t_1}{8}(3\beta_1^2+2\beta_1\beta_2+3\beta_2^2)+\dfrac{t_2}{16}
 \left(5\beta_1^3+
  3\beta_1^2\beta_2+3\beta_1\beta_2^2+5\beta_2^3\right) \nonumber \\
  &+
  \dfrac{t_3}{128}(35\beta_1^4+20\beta_1^3\beta_2+18\beta_1^2\beta_2^2+20\beta_1\beta_2^3+35\beta_2^4)+\cdots.
  \end{align}
 The hodograph equations \eqref{crit} with $t_{n}\,=\,0$ for $n\,\geq\,4$ take the form
 \begin{equation}\label{h2}
\everymath{\displaystyle}
\begin{cases}
8x+4t_{1}(3\beta_1+\beta_2)+3t_{2}\left(5\beta_1^2+
  2\beta_1\beta_2+\beta_2^2\right)+
  \dfrac{t_3}{8}(140\beta_1^3+60\beta_1^2\beta_2+36\beta_1\beta_2^2+20\beta_2^2)=0,\\\\
8x+4t_1(\beta_1+3\beta_2)+3t_2\left(\beta_1^2+
  2\beta_1\beta_2+5\beta_2^2\right)
  +\dfrac{t_3}{8}(140\beta_2^3+60\beta_2^2\beta_1+36\beta_2\beta_1^2+20\beta_1^2)=0.
\end{cases}
\end{equation}
Detailed analysis of equations \eqref{h2} will be performed in section 4.

\vspace{0.3cm}
\subsection*{dToda hierarchy}

It is known (see e.g. \cite{gin}) that as $N\rightarrow\infty$ the asymptotic density of  eigenvalues $\{\lambda_1,\ldots,\lambda_N\}$ of the $H$ model in the zero genus case  concentrates on a single interval $[\beta_1,\beta_2]$  and is given by  \cite{gin}
\[
\rho(\lambda,\bt,\bb):=\dfrac{h_H(\lambda,\bt,\bb)}{2\,\pi\,i\,T}\,\sqrt{(\lambda-\beta_1)(\lambda-\beta_2)},
\]
where
\[
h_H(\lambda,\bt,\bb):=\Big(\dfrac{V_{H,\lambda}(\lambda,\bt)}{\sqrt{(\lambda-\beta_1)(\lambda-\beta_2)}}\Big)_{\oplus}.
\]
Here $\oplus$ denotes the projection on strictly positive powers of $\lambda$. Moreover, the
 endpoints $\beta_1$ and $\beta_2$  of the eigenvalue support are determined by the equations
\begin{equation}\label{rm}
\oint_{\gamma}\dfrac{\d \lambda}{2\,i\,\pi}\,\dfrac{\partial_{\lambda}V_{H}(\lambda,\bt)}{\sqrt{
(\lambda-\beta_1)\,(\lambda-\beta_2)}}=0,\quad
\oint_{\gamma}\dfrac{\d \lambda}{2\,i\,\pi}\,\dfrac{\lambda\,\partial_{\lambda}V_{H}(\lambda,\bt)}{\sqrt{
(\lambda-\beta_1)\,(\lambda-\beta_2)}}=2\,T,
\end{equation}
where
\[
V_H(\lambda,\bt):=\sum_{n\geq 1} \lambda^n\,t_{n},
\]
and $\gamma$ denotes a large positively oriented circle $|\lambda|=R$.
It is immediate to see   that these equations are equivalent to the system \eqref{crit} for  the critical points of the function
\begin{equation}\label{w2}
W_H(\bt,\bb)\,:=T\,(\beta_1+\beta_2)+\oint_{\gamma}\dfrac{\d \lambda}{2\,i\,\pi}\,\sqrt{(\lambda-\beta_1)\,(\lambda-\beta_2)
}\,\partial_{\lambda}V_{H}(\lambda,\bt),
\end{equation}
which satisfies the Euler-Poisson-Darboux equation $E(-\frac{1}{2}, -\frac{1}{2})$. It is also easy to check that  the functions $W_B$ and $W_H$ are related according to
\begin{equation}\label{trans}
W_B=4\,\dfrac{\partial^2
W_H}{\partial\beta_1\,\partial\beta_2}-t_{H,1},
\quad
 t_{B,n}=(n+2)\,t_{H,n+2},\quad n\geq 0.
\end{equation}

The first terms of the function $W_H$ are
\begin{align*}\everymath{\displaystyle}
W_H&=T (\beta_1+\beta_2)-\frac{t_1}{8}(\beta_1-\beta_2)^2-\frac{t_{2}}{8}(\beta_1+\beta_2) (\beta_1-\beta_2)^2
 -\frac{3t_{3}}{128}\left(5 \beta_1^2+6 \beta_2\beta_1+5 \beta_2^2\right) (\beta_1-\beta_2)^2\\
 & -\frac{t_{4}}{64}\left(7 \beta_1^3+9 \beta_2\beta_1^2+9 \beta_2^2 \beta_1+7 \beta_2^3\right) (\beta_1-
   \beta_2)^2+\cdots,
\end{align*}
The corresponding equations \eqref{crit} for $t_{n}=0,\; \forall n\geq 4$ take the form
$$\everymath{\displaystyle}\begin{cases}
T+\frac{t_1}{4}(\beta_2-\beta_1)+\frac{t_{2}}{8}\left(-3\beta_1^2+2 \beta_2 \beta_1+\beta_2^2\right)-\frac{3t_{3}}{32}
   \left(5 \beta_1^3-3 \beta_2 \beta_1^2-\beta_2^2 \beta_1-
   \beta_2^3\right)=0,\\  \\
T+\frac{t_1}{4}(\beta_1-\beta_2)+\frac{t_{2}}{8}\left(\beta_1^2+2 \beta_2 \beta_1-3 \beta_2^2\right)+\frac{3t_{3}}{32}
   \left(\beta_1^3+\beta_2   \beta_1^2+3 \beta_2^2 \beta_1-5 \beta_2^3\right)=0.
\end{cases}$$

It is known that the  hodograph system \eqref{rm}  provides
solutions of the dispersionless Toda hierarchy (see e.g.
\cite{se,dsl}). Indeed,  for "times" $T$ and $t_1$  one gets
\begin{equation}\label{todr}
\begin{cases}
\partial_{t_1}\,\beta_1=-\dfrac{1}{4}\,(\beta_1-\beta_2)\,\beta_{1T},\\ \\
\partial_{t_1}\,\beta_2=\dfrac{1}{4}\,(\beta_1-\beta_2)\,\beta_{2T},
\end{cases}
\end{equation}
which in terms of  variables
\begin{equation}\label{tod}
u=\dfrac{1}{2}\,(\beta_1+\beta_2), \quad v=\dfrac{1}{16}\,(\beta_1-\beta_2)^2,
\end{equation}
becomes the dToda system
\eqref{dToda}.  Moreover, for times $t_1,t_2$ the hodograph equations  \eqref{rm}  imply the Benney
system \eqref{ben}.
If we use the dependent variables $(u,v)$ and introduce the function
\begin{equation}\label{d1}
f(\bt,u,v):=W_H(\bt,u,v)-2\,T\,u+2\,t_1\,v+4\,t_2\,u\,v,
\end{equation}
then the hodograph equations read
\begin{equation}\label{d2}
\begin{cases}
2\,T-4\,v\,t_2+\partial_u\,f=0,\\
-2\,t_1-4\,u\,t_2+\partial_v\,f=0,
\end{cases}
\end{equation}
which, under trivial rescalings and the substitution
$(u,v)\mapsto(v,u)$,  coincides with the hyperbolic version of
hodograph system (2.11) of \cite{dubgr}. In particular, from the
EPD equation \eqref{edpa} it follows easily that
\begin{equation}\label{d3}
\dfrac{\partial^2\, W_H}{\partial\, u^2}-v\,\dfrac{\partial^2\, W_H}{\partial\, v^2}=\dfrac{\partial^2\, f}{\partial\, u^2}-v\,\dfrac{\partial^2\, f}{\partial\, v^2}=0.
\end{equation}

\section{Characterization of singular sectors}

 Using \eqref{crit} and \eqref{edpa} we may now analyze the structure of singular sectors of the $B$ system and the $H$ model in a unified way. Thus, let us  denote by $\mathcal{M}$ the set of solutions $(\bt,\bb)\, (\beta_1\neq \beta_2)$ of the hodograph equations  \eqref{crit}. There is a partition of $\mathcal{M}$ into  a  regular and a singular sector
\[
\mathcal{M}={\mathcal{M}}^{\mbox{reg}}\cup{\mathcal{M}}^{\mbox{sing}},
\]
such that given $(\bt,\bb)\in\mathcal{M}$
\[
\mbox{$(\bt,\bb)\in\mathcal{M}^{\mbox{reg}}$ if $\det \Big(\dfrac{\partial^2\,W(\bt,\bb)}{\partial\,\beta_i\,\partial\,\beta_j}\Big)\neq 0$},\quad
\mbox{$(\bt,\bb)\in\mathcal{M}^{\mbox{sing}}$ if $\det \Big(\dfrac{\partial^2\,W(\bt,\bb)}{\partial\,\beta_i\,\partial\,\beta_j}\Big)=0$}.
\]
 The elements of $\mathcal{M}^{\mbox{reg}}$, correspond to the case when the system \eqref{crit} is uniquely solvable and hence, it defines a unique solution $\bb(\bt)$.
The singular sector  $\mathcal{M}^{\mbox{sing}}$  contains the degenerate critical points of the function $W$ on  which the implicit solutions $\bb(\bt)$ of the  hodograph equations exhibit  \emph{gradient catastrophe}
behaviour.

The Euler-Poisson-Darboux equation is of great help to analyze  the structure of  $\mathcal{M}^{\mbox{sing}}$. Indeed, if $(\bt,\bb)\in \mathcal{M}$,  as a consequence of \eqref{edpa} it is clear that
 \begin{equation}\label{nondiag}
\frac{\partial^2\,W}{\partial\,\beta_1\,\partial\,\beta_2}\,=\,0.
\end{equation}
Consequently
\begin{equation}\label{det}
\det\Big(\frac{\partial^2\,W}{\partial\,\beta_i\,\partial\,\beta_j}\Big)=
\frac{\partial^2\,W}{\partial\,\beta_1^2}\cdot\frac{\partial^2\,W}{\partial\,\beta_2^2}.
\end{equation}
Thus, we have
\begin{pro}
Given $(\bt,\bb)\in \mathcal{M}$ then
\begin{enumerate}
\item $(\bt,\bb)\,\in\,\mathcal{M}^{\mbox{reg}}$ if and only if
$\everymath{\displaystyle}
\frac{\partial^2\,W}{\partial\,\beta_1^2}\,\neq\,0\, \mbox{and} \,
\frac{\partial^2\,W}{\partial\,\beta_2^2}\,\neq\,0.
$
\item $(\bt,\bb)\,\in\,\mathcal{M}^{\mbox{sing}}$ if and only at least one of the
derivatives
$\everymath{\displaystyle}
\frac{\partial^2\,W}{\partial\,\beta_1^2},\, \frac{\partial^2\,W}{\partial\,\beta_2^2},
$
vanishes.
\end{enumerate}
\end{pro}
Furthermore, using \eqref{edpa} it follows at once  that   at any point $(\bt,\bb)\in {\mathcal{M}}$ all mixed derivatives $\partial_{\beta_1}^i\partial_{\beta_2}^j W$ can be expressed in terms of linear combination of \emph{diagonal} derivatives  $\partial_{\beta_1}^nW$ and  $\partial_{\beta_2}^m W$. Let us now define ${\mathcal{M}}^{\mbox{sing}}_{n_1,n_2}$ as the set of points $(\bt,\bb)\in {\mathcal{M}}$ such that
\begin{equation}\label{m12}
\dfrac{\partial^{n_i+2}W}{\partial\beta_i^{n_i+2}}\neq 0; \quad
\dfrac{\partial^k W}{\partial \beta_i^k}=0,\quad  \forall 1\leq k\leq n_i+1,\quad (i=1,2).
\end{equation}
It is clear that
$$
{\mathcal{M}}^{\mbox{sing}}_{n_1,n_2}\bigcap{\mathcal{M}}^{\mbox{sing}}_{n'_1,n'_2}=\emptyset,\;
\mbox{ for $(n_1,n_2)\neq (n'_1,n'_2)$}
$$
Then, it follows at once that
\begin{pro}
The singular sector satisfies
\begin{equation}\label{singp}
{\mathcal{M}}^{\mbox{sing}}=\bigcup _{n_1+n_2 \geq 1}{\mathcal{M}}^{\mbox{sing}}_{n_1,n_2}.
\end{equation}
\end{pro}

 According to \eqref{trans} and using \eqref{edpa} we have
\[
W_B=-t_{H,1}-\dfrac{2}{\beta_1-\beta_2}\Big(\dfrac{\partial \,W_H}{\partial \beta_1}-\dfrac{\partial \,W_H}{\partial \beta_2}\Big),\quad t_{B,n}=(n+2)\,t_{H,n+2},\quad n\geq 0.
\]
It is then immediate to conclude that
\begin{pro} The singular sectors of the $H$ model and the Benney system, with  \newline $ t_{B,n}=(n+2)\,t_{H,n+2}$ for $n\geq 0$, satisfy
\begin{equation}\label{rel}
\begin{cases}
\mathcal{M}^{\mbox{sing}}_{H,1,1}\subset \mathcal{M}^{\mbox{reg}}_B,\\\\
\mathcal{M}^{\mbox{sing}}_{H,n_1+1,n_2+1}\subset \mathcal{M}^{\mbox{sing}}_{B,n_1,n_2} \; \mbox{for $n_1\geq 1,n_2\geq 1$}.
\end{cases}
\end{equation}

\end{pro}
\subsection*{Singular sectors in the $H$ model}

In the applications of the $H$ model in quantum gravity \cite{gin} the singular sectors are described in terms of the behaviour of the eigenvalue density $\rho(\lambda)$
at the endpoints $(\beta_1,\beta_2)$ of its support. We will now show how this description derives in a natural way from
the classification \eqref{singp}
based on the subsets  $\mathcal{M}^{\mbox{sing}}_{H,n_1,n_2}$ of the singular sector . To this end let us consider the derivative $\partial_{\beta_1}^{k+1}W_H$ with $k\geq 2$ , it is obviously proportional to the integral
\begin{equation*}\everymath{\displaystyle}
\oint_{\gamma}\dfrac{\d \lambda}{2\,i\,\pi}\
(\lambda-\beta_2)\,\dfrac{V_{H,\lambda}/\sqrt{(\lambda-\beta_1)\,(\lambda-\beta_2)}}{
(\lambda-\beta_1)^k}=\oint_{\gamma}\dfrac{\d \lambda}{2\,i\,\pi}\
\dfrac{(\lambda-\beta_2)\,h_H(\lambda)}{
(\lambda-\beta_1)^k}
=\dfrac{\partial_{\lambda}^{k-1}}{(k-1)!}\, \Big((\lambda-\beta_2)\,h_H(\lambda)\Big)\Big|_{\lambda=\beta_1},
\end{equation*}
and a similar result follows for the derivatives  $\partial_{\beta_2}^{k+1}W$ with $k\geq 2$.  As a consequence we have
\begin{pro}
A point  $(\bt,\bb)\in\mathcal{M}_H$ belongs to the singularity class $\mathcal{M}^{\mbox{sing}}_{H,n_1,n_2}$ if and only if
\begin{equation}
\everymath{\displaystyle}
\rho(\lambda,\bt,\bb)\sim (\lambda-\beta_i)^{\frac{2n_i+1}{2}}\quad \mbox{as $\lambda\rightarrow \beta_i,\quad (i=1,2)$}
\end{equation}
\end{pro}
\noindent
This property is crucial to establish the relationship between the regularized singularity sectors of the $H$ model and the minimal conformal models $(p,q)$ with $q=2$ \cite{gin}.

\subsection*{Singular sectors in the $B$ system}

\vspace{0.3cm}

 In analogy with the $H$ model we may  characterize  the  classes $\mathcal{M}^{\mbox{sing}}_{B,n_1,n_2}$ of the singular sector of the $B$ system in terms of the behaviour of $S(\lambda)$
at  $\lambda=\beta_i\, (i=1,2)$ . Indeed the derivative $\partial_{\beta_1}^{k+1}W_B$ with $k\geq 1$ is proportional to the integral
\begin{equation*}\everymath{\displaystyle}
\oint_{\gamma}\dfrac{\d \lambda}{2\,i\,\pi}\
(\lambda-\beta_2)\,\dfrac{\lambda\,V_B/\sqrt{(\lambda-\beta_1)\,(\lambda-\beta_2)}}{
(\lambda-\beta_1)^{k+1}}=\oint_{\gamma}\dfrac{\d \lambda}{2\,i\,\pi}\
\dfrac{(\lambda-\beta_2)\,h_B(\lambda)}{
(\lambda-\beta_1)^{k+1}}
=\dfrac{\partial_{\lambda}^{k} }{k!}\,\Big((\lambda-\beta_2)\,h_B(\lambda)\Big)\Big|_{\lambda=\beta_1},
\end{equation*}
and a similar result follows for the derivatives  $\partial_{\beta_2}^{k+1}W_B$ with $k\geq 2$.  As a consequence we have
\begin{pro}
A point  $(\bt,\bb)\in\mathcal{M}_B$ belongs to the singularity class $\mathcal{M}^{\mbox{sing}}_{B,n_1,n_2}$ if and only if
\begin{equation}
\everymath{\displaystyle}
S(\lambda,\bt,\bb)\sim (\lambda-\beta_i)^{\frac{2n_i+3}{2}}\quad \mbox{as $\lambda\rightarrow \beta_i,\quad (i=1,2)$}
\end{equation}
\end{pro}
This result shows the existing duality  between the eigenvalue density  $\rho$ of the $H$ model and the $S$ function of
the $B$ system.

\section{Explicit determination of singular sectors  in terms of constrains $f_k(\bt)=0$}

The singular classes  ${\mathcal{M}}^{\mbox{sing}}_{n_1,n_2}$ can be determined by means of systems of $n_1+n_2$ constraints for the coordinates $\bt$.  To see this property notice that the points $(\bt,\bb)$ of  $ {\mathcal{M}}^{\mbox{sing}}_{n_1,n_2}$ are characterized by the equations
\begin{equation}\label{m12b}
\dfrac{\partial^k W}{\partial \beta_i^k}=0,\quad  \forall 1\leq k\leq n_i+1,\quad i=1,2,
\end{equation}
and
\begin{equation}\label{m12c}
\dfrac{\partial^{n_i+2}W}{\partial\beta_i^{n_i+2}}\neq 0, \quad i=1,2.
\end{equation}
Furthermore,  the jacobian matrix of the the system of two equations
\begin{equation}\label{m12d}
\dfrac{\partial^{n_i+1}W}{\partial\beta_i^{n_i+1}}=0, \quad i=1,2
\end{equation}
is not singular as
\begin{equation}\label{Delta23}\everymath{\displaystyle}
\Delta\,:=\,\left|\begin{array}{cc}
\frac{\partial^{n_1+2}\,W}{\partial\,\beta_1^{n_1+2}} & \frac{\partial^{n_2+2}\,W}{\partial\, \beta_1\partial\,\beta_2^{n_2+1}}
 \\  \\
 \frac{\partial^{n_1+2}\,W}{\partial\,\beta_1^{n_1+1}\,\partial\,\beta_2} & \frac{\partial^{n_2+2}\,W}{\partial\,\beta_2^{n_2+2}}
\end{array}\right|\,\neq\,0.
\end{equation}
Indeed, we notice that as a consequence of \eqref{edpa} the derivatives outside the diagonal of $\Delta$ are  linear combinations of  the derivatives  $\{\partial_{\beta_i}^k\,W, \; 1\leq k\leq n_i+1,\; i=1,2\}$, so that from \eqref{m12b}-\eqref{m12c} we have
\[
\Delta=\frac{\partial^{n_1+2}\,W}{\partial\,\beta_1^{n_1+2}}\cdot \frac{\partial^{n_2+2}\,W}{\partial\,\beta_2^{n_2+2}}\neq 0.
\]
Therefore, one can solve \eqref{m12d} and get a solution $\bb(\bt)$. Substituting this solution in the remaining equations
\eqref{m12b} gives $n_1+n_2$ constraints of the form
\[
f_k(\bt)=0,\quad k=1,\ldots,n_1+n_2.
\]
It is not difficult to determine the solutions  of \eqref{m12b}-\eqref{m12c}  when $t_n=0,\, n\geq 4$. Lists of such solutions are given next.

\subsection*{Example:  the $H$ model }

Let us consider the $H$ model and assume that  $t_n=0$ if $n\geq4$ . Then it follows  that the condition $\det (\partial_{\beta_i\beta_j}\,W_H(\bt,\bb))=0$ which characterizes ${\mathcal{M}}^{\mbox{sing}}_H$ reads
$$
32t_2^2-24xt_3+72t_2t_3(\beta_1+\beta_2)+27t_3^2\left(\beta_1^2+6\beta_2\beta_1+\beta_2^2\right)=0.
$$
There are two generic classes in ${\mathcal{M}}^{\mbox{sing}}_H$ given by ${\mathcal{M}}^{\mbox{sing}}_{H,1,0}$ and ${\mathcal{M}}^{\mbox{sing}}_{H,0,1}$. Assuming that $(t_1,t_2,t_3,\beta_1,\beta_2)$ must be real,  both  classes  include  two  cases

$$\everymath{\displaystyle}
{\mathcal{M}}^{\mbox{sing}}_{H,1,0}:
\begin{cases}
1)\quad  x=\frac{2 t_2^2-9 \sqrt[3]{6T^2}\,t_3^{4/3}}{6 t_3},\\�\\
\quad \beta_1=-\frac{2 t_2+\sqrt[3]{36T}\,t_3^{2/3}}{6 t_3},\quad
\beta_2=\frac{-2t_2+3\sqrt[3]{36T}\,t_3^{2/3}}{6 t_3}.\\  \\  \\
2)\quad  x=\frac{4 t_2^2+9 \sqrt[3]{6T^2}\,\left(1-i
\sqrt{3}\right)t_3^{4/3}}{12 t_3},\\  \\
 \quad  \beta_1=\frac{-4t_2+\left(1+i\sqrt{3}\right)\sqrt[3]{36T}\, t_3^{2/3}}{12 t_3},\quad
\beta_2=\frac{-4 t_2-3\left(1+i\sqrt{3}\right)\sqrt[3]{36T}\,t_3^{2/3}}{12 t_3},
\end{cases}$$
where we assume $t_3<0$ in the second case.

$$\everymath{\displaystyle}
{\mathcal{M}}^{\mbox{sing}}_{H,0,1}\begin{cases}
1)\quad  x=\frac{2t_2^2-9 \sqrt[3]{6T^2}\,t_3^{4/3}}{6 t_3},\\ \\
\quad \beta_1=\frac{-2t_2+3\sqrt[3]{36T}t_3^{2/3}}{6 t_3},\quad
\beta_2=-\frac{2 t_2+\sqrt[3]{36T}t_3^{2/3}}{6 t_3},\\  \\  \\
2) \quad   x=\frac{4 t_2^2+9 \sqrt[3]{6T^2} \left(1-i
\sqrt{3}\right)t_3^{4/3}}{12 t_3},\\  \\
  \quad \beta_1=\frac{-4 t_2-3\left(1+i\sqrt{3}\right)\sqrt[3]{36T}t_3^{2/3}}{12 t_3},\quad
\beta_2=\frac{-4t_2+\left(1+i\sqrt{3}\right)\sqrt[3]{36T}\, t_3^{2/3}}{12 t_3},
\end{cases}$$
where we assume $t_3<0$ in the second case.


\subsection*{Example: the $B$ system}

Let us now consider the system of hodograph equations for
the $B$ system with $t_n=0$ for all $n\geq4$. Now the condition $\mbox{det} (\partial_{\beta_i\beta_j}
W_B(\bt,\bb))=0$ reduces to
$$\everymath{\displaystyle}\begin{array}{l}
32 t_1^2+96 t_2 (\beta_1+\beta_2) +702 t_3^2 \beta_1^2 \beta_2^2+72
   \left(3 t_2^2+t_1 t_3\right) \beta_1
   \beta_2+12 \left(3 t_2^2+13 t_1 t_3\right)
   \left(\beta_1^2+\beta_2^2\right)+\\  \\
   \quad 486 t_2 t_3   \left(\beta_2 \beta_1^2+\beta_2^2 \beta_1\right)+90 t_2 t_3 \left(\beta_1^3+
   \beta_2^3\right)+180 t_3^2 \left(\beta_2 \beta_1^3+\beta_2^3 \beta_1\right)+45 t_3^2
   \left(\beta_1^4+\beta_2^4\right)=0.
\end{array}$$
There are two classes in $\mathcal{M}^{\mbox{sing}}_{B,1,0}$

$$\everymath{\displaystyle}\begin{array}{ll}
\textbf{1.} & x\,=\,\frac{-45t_3 t_2^3+180 t_1 t_3^2
   t_2+\sqrt{15}(8t_1t_3-3t_2^2)\sqrt{t_3^2 \left(3 t_2^2-8
   t_1 t_3\right)}}{360 t_3^3},\\  \\
   &\beta_1\,=\,-\frac{5 t_2 t_3+\sqrt{15} \sqrt{t_3^2 \left(3
   t_2^2-8 t_1 t_3\right)}}{20t_3^2},\qquad \beta_2\,=\,\frac{-3 t_2 t_3+\sqrt{15} \sqrt{t_3^2 \left(3 t_2^2-8 t_1
   t_3\right)}}{12 t_3^2},\\  \\  \\
\textbf{2.} & x\,=\,\frac{-45t_3 t_2^3+180 t_1 t_3^2
   t_2-\sqrt{15}(8t_1t_3-3t_2^2)\sqrt{t_3^2 \left(3 t_2^2-8
   t_1t_3\right)}}{360 t_3^3},\\  \\
   &\beta_1\,=\,\frac{-5 t_2 t_3+\sqrt{15} \sqrt{t_3^2 \left(3
   t_2^2-8 t_1 t_3\right)}}{20t_3^2},\qquad \beta_2\,=\,-\frac{3 t_2 t_3+\sqrt{15} \sqrt{t_3^2 \left(3 t_2^2-8 t_1
   t_3\right)}}{12 t_3^2}
     \end{array}$$
In the same way it follows that  $\mathcal{M}^{\mbox{sing}}_{B,0,1}$ has two classes given by
$$\everymath{\displaystyle}\begin{array}{ll}
\textbf{1.} & x\,=\,\frac{-45t_3t_2^3+180 t_1 t_3^2
   t_2-\sqrt{15}(8t_1t_3-3t_2^2)\sqrt{t_3^2 \left(3 t_2^2-8
   t_1t_3\right)}}{360 t_3^3},\\  \\
   &\beta_1\,=\,-\frac{3 t_2 t_3+\sqrt{15} \sqrt{t_3^2 \left(3 t_2^2-8 t_1
   t_3\right)}}{12 t_3^2},\qquad \beta_2\,=\,\frac{-5 t_2 t_3+\sqrt{15} \sqrt{t_3^2 \left(3
   t_2^2-8 t_1 t_3\right)}}{20t_3^2},\\  \\  \\
\textbf{2.} & x\,=\,\frac{-45t_3t_2^3+180 t_1t_3^2
   t_2+\sqrt{15}(8t_1t_3-3t_2^2)\sqrt{t_3^2 \left(3 t_2^2-8
   t_1 t_3\right)}}{360 t_3^3},\\  \\
   &\beta_1\,=\,\frac{-3 t_2 t_3+\sqrt{15} \sqrt{t_3^2 \left(3 t_2^2-8 t_1
   t_3\right)}}{12 t_3^2},\qquad \beta_2\,=\,-\frac{5 t_2 t_3+\sqrt{15} \sqrt{t_3^2 \left(3
   t_2^2-8 t_1 t_3\right)}}{20t_3^2}.
   \end{array}$$

\vspace{0.5cm}

\subsection* { Acknowledgements}

\vspace{0.3cm} The authors  wish to thank the  Spanish Ministerio de
Educaci\'on y Ciencia (research project FIS2008-00200/FIS) for its
finantial support. B. K. is thankful to the Departamento de F\'{i}sica Te\'orica II for the kind hospitality.

\end{document}